\documentclass[12pt]{article}

\RequirePackage{etex}

\usepackage[margin=1in]{geometry}   
\usepackage{amsmath,amsfonts,amssymb,amsthm}
\usepackage{booktabs}
\usepackage{multirow}
\usepackage[onehalfspacing]{setspace} 
\renewcommand{\arraystretch}{0.6} 

\usepackage{float}
\usepackage{caption}
\usepackage{graphicx}
\usepackage{adjustbox}
\usepackage{lscape}

\usepackage{adjustbox}

\usepackage{xcolor}

\usepackage{relsize}

\usepackage{bbm}
\usepackage{subfig}

\usepackage{authblk}

\usepackage{amsthm}

\newtheorem*{proposition*}{Proposition}

\usepackage{xpatch}
\makeatletter
\xpatchcmd{\proof}{\@addpunct{.}}{\@addpunct{:}}{}{}
\makeatother

\makeatletter
\newcommand{\vast}{\bBigg@{3}}
\newcommand{\Vast}{\bBigg@{4}}
\makeatother

\newcommand\independent{\protect\mathpalette{\protect\independenT}{\perp}}
\def\independenT#1#2{\mathrel{\rlap{$#1#2$}\mkern2mu{#1#2}}}

\makeatletter
\newcommand*{\indep}{%
  \mathbin{%
    \mathpalette{\@indep}{}%
  }%
}
\newcommand*{\nindep}{%
  \mathbin{
    \mathpalette{\@indep}{\not}
  }%
}
\newcommand*{\@indep}[2]{%
  \sbox0{$#1\perp\m@th$}
  \sbox2{$#1=$}
  \sbox4{$#1\vcenter{}$}
  \rlap{\copy0}
  \dimen@=\dimexpr\ht2-\ht4-.2pt\relax
  \kern\dimen@
  {#2}%
  \kern\dimen@
  \copy0 
} 
\makeatother

\DeclareMathOperator{\E}{\textnormal{\mbox{E}}}

\usepackage[utf8]{inputenc}
\DeclareUnicodeCharacter{200E}{}

\usepackage{setspace}
\usepackage{cite}

\usepackage{xr-hyper}

\usepackage{xcolor}
\definecolor{forestgreen}{RGB}{34,139,34}
\usepackage{hyperref}
\hypersetup{
    colorlinks=true,
    linkcolor=magenta,
    filecolor=magenta, 
    citecolor=forestgreen,      
    urlcolor=blue
}

\makeatletter
\newcommand*{\addFileDependency}[1]{
  \typeout{(#1)}
  \@addtofilelist{#1}
  \IfFileExists{#1}{}{\typeout{No file #1.}}
}
\makeatother

\newcommand*{\myexternaldocument}[1]{%
    \externaldocument{#1}%
    \addFileDependency{#1.tex}%
    \addFileDependency{#1.aux}%
}

\myexternaldocument{02_appendix}

\usepackage[stable]{footmisc}

\usepackage{rotating}

\makeatletter
\def\@hangfrom#1{\setbox\@tempboxa\hbox{{#1}}%
      \hangindent 0pt
      \noindent\box\@tempboxa}
\makeatother

\makeatletter
\def\@seccntformat#1{\@ifundefined{#1@cntformat}%
   {\csname the#1\endcsname\quad}  
   {\csname #1@cntformat\endcsname}
}
\let\oldappendix\appendix 
\renewcommand\appendix{%
    \oldappendix
    \newcommand{\section@cntformat}{\appendixname~\thesection\quad}
}
\makeatother

\usepackage[absolute,showboxes]{textpos}

\TPMargin{5pt}

\newcommand{\copyrightstatement}{
    \begin{textblock}{0.84}(0.08,0.93)    
         \noindent
         \footnotesize
         This draft manuscript presents work-in-progress. Comments are welcome at \href{mailto:idahabreh@hsph.harvard.edu}{idahabreh@hsph.harvard.edu}.  \\
         The Appendix is available at this \href{https://www.dropbox.com/s/vsinnebqgpscipl/benchmarking_and_joint_inference_appendix.pdf?dl=0}{link}. 
    \end{textblock}
}

\usepackage{datetime}
\def\paperversionmajor{12}
\def\paperversionminor{0}

\usepackage{sectsty}
\subsectionfont{\noindent\textbf\itshape}
\subsubsectionfont{\normalfont\itshape}

\begin{document}


\title{Randomized trials and their observational emulations: a framework for benchmarking and joint analysis \vspace*{0.3in} }

\copyrightstatement

\author[1-3]{Issa J. Dahabreh 
}
\author[4]{Jon A. Steingrimsson}
\author[1-3]{James M. Robins}
\author[1-3,5]{Miguel A. Hern\'an}

\affil[1]{CAUSALab, Harvard T.H. Chan School of Public Health, Boston, MA USA}
\affil[2]{Department of Epidemiology, Harvard T.H. Chan School of Public Health, Boston, MA}
\affil[3]{Department of Biostatistics, Harvard T.H. Chan School of Public Health, Boston, MA}
\affil[4]{Department of Biostatistics, School of Public Health, Brown University, Providence, RI}
\affil[5]{Harvard-MIT Division of Health Sciences and Technology, Boston, MA}

\maketitle{}

\thispagestyle{empty}

\clearpage 

\vspace{0.3in}
\noindent
\textbf{Running head:} Framework for benchmarking and joint analysis

\vspace{0.3in}
\noindent
\textbf{Type of manuscript:} Original Research Article.

\vspace{0.3in}
\noindent
\textbf{Conflicts of interest:} None declared.

\vspace{0.3in}
\noindent
\textbf{Data and computing code availability:} Not applicable.

\vspace{0.3in}
\noindent
\textbf{Word count:} abstract = 248; main text $\approx$ 4000.

\vspace{0.3in}
\noindent
\textbf{Abbreviations that appear in the text:} No abbreviations.

\thispagestyle{empty}

\clearpage
\thispagestyle{empty}

\vspace*{1in}

\begin{abstract}
\noindent
\linespread{1.4}\selectfont
A randomized trial and an analysis of observational data designed to emulate the trial sample observations separately, but have the same eligibility criteria, collect information on some shared baseline covariates, and compare the effects of the same treatments on the same outcomes. Treatment effect estimates from the trial and its emulation can be compared to benchmark observational analysis methods. In a simplified setting with complete adherence to the assigned treatment strategy and no loss-to-follow-up, we show that benchmarking relies on an exchangeability condition between the populations underlying the trial and its emulation, to account for differences in the distribution of covariates between them. When this exchangeability condition holds, and the usual conditions needed for the estimates from the trial and its emulation to have a causal interpretation also hold, we derive restrictions on the law of the observed data. When the data are compatible with the restrictions, joint analysis of the trial and its emulation is possible. When the data are incompatible with the restrictions, a discrepancy between (1) estimates based on extending inferences from the trial to the population underlying the emulation and (2) the emulation itself may reflect either inability to benchmark (e.g., due to selective participation into the trial) or a failure of the emulation (e.g., due to unmeasured confounding), but we cannot use the data to determine which is the case. Our analysis reveals how benchmarking attempts combine causal assumptions, data analysis methods, and substantive knowledge to examine the validity of observational analysis methods. \\

\vspace{0.2in}
\noindent
\textbf{Keywords:} randomized trials; observational analyses; target trial emulation; transportability; benchmarking.

\end{abstract}

\clearpage
\setcounter{page}{1}
\section*{Introduction}

\noindent
Consider an index randomized trial and an analysis of observational data designed to emulate a target trial as similar as possible to the index trial \cite{hernan2016}. The trial and the observational emulation sample observations separately, but have the same eligibility criteria, collect information on some shared baseline covariates, compare the same treatment recommendations (assigned randomly in the trial but not in the observational emulation), and examine the same outcomes \cite{dahabreh2020studydesigns, dahabreh2020benchmarking}. Because of the commonalities of the two studies, it is natural to want to compare their results and jointly use their data to learn about the effects of the intervention.

Causal inference from observational data is typically considered more speculative than causal inference in trials due to the possibility of baseline confounding by unmeasured variables in the absence of randomization. Agreement between the results of a trial and its emulation might then be taken to indicate that the methods in the emulation have in some sense been ``successful'' in producing valid causal inferences \cite{dahabreh2014can, hernan2016}. We refer to the comparison of the results of the trial and the emulation as benchmarking the observational analyses \cite{dahabreh2020benchmarking}. By benchmarking in cases where data are conveniently available from pragmatic trials, we can develop confidence in the reliability of observational methods when used in settings where no trial data are available \cite{hernan2016}. For example, if several benchmarking attempts in some clinical domain suggest that observational analyses  successfully emulate the results of large well-run pragmatic trials, we might be willing to trust the results from similar observational analyses for treatments that have not been investigated in trials, or for follow-up times and patient subgroups for which the trial results are inadequate (e.g., due to short follow-up duration or limited sample size) \cite{forbes2020benchmarking}.

Here, we examine a simplified setting with complete adherence to treatment and no loss to follow-up to show that benchmarking relies on an exchangeability assumption between the populations underlying the trial and its emulation to account for differences in the distribution of covariates between them and to allow transportability of inferences from the trial to the population underlying the emulation. We show that, when this exchangeability condition holds, and the ``usual'' conditions needed for treatment effect estimates from the trial and its emulation to have a causal interpretation also hold \cite{hernan2020}, they imply restrictions for the observed data distribution, allowing investigators to examine whether the data are compatible with the assumptions. When the data are incompatible with these restrictions, a discrepancy between (1) estimates based on extending inferences from the trial to the population underlying the emulation and (2) the emulation itself may reflect either inability to benchmark (e.g., due to selective participation into the trial) or a failure of the emulation (e.g., due to unmeasured confounding), but we cannot use the data to decide which is the case. We also show that if the exchangeability conditions hold, joint analysis to estimate potential outcome means and treatment effects is feasible and can be more efficient than either relying on just the observational analysis or the transportability analysis from the trial to population represented by the emulation.

\section*{Setup and causal estimands}

Suppose that a trial and an analysis of observational data designed to emulate the trial measure some shared baseline (pre-treatment) covariates $X$, compare the same treatment strategies $A$ (randomly assigned in the trial but not in the emulation), and examine the same outcome $Y$, measured at a single post-treatment time-point \cite{dahabreh2020studydesigns}. We use $S$ as the binary indicator for participation status ($S = 1$ for the trial and $S = 0$ for the emulation). 

The trial's eligibility criteria define the actual (finite) population of all trial-eligible individuals. We view the actual population as a simple random sample from a near-infinite underlying super-population. Typically, the actual population is not enumerated and thus the individuals who participate in the trial represent an unknown fraction of all trial-eligible individuals \cite{dahabreh2020studydesigns}. For simplicity, we assume that the observational emulation is conducted among a simple random sample of trial-eligible non-randomized individuals from the actual population, with unknown sampling probability (this is the sampling scheme of non-nested trial designs \cite{dahabreh2020studydesigns, dahabreh2019generalizing}). The simple random sampling assumption can be relaxed when at least partial information is available about the selection of non-randomized individuals \cite{dahabreh2020studydesigns, dahabreh2019generalizing}; we do not consider such situations to maintain focus on concepts related to benchmarking.

The data from the trial are independent realizations of $(X_i, S_i = 1, A_i, Y_i)$, $i = 1, \ldots, n_{1}$, where $n_1$ is the total number of trial participants; the data from the emulation are independent realizations of $(X_i, S_i = 0, A_i, Y_i)$, $i = 1, \ldots, n_{0}$, where $n_0$ is the total number of participants in the emulation. The total sample size is $n = n_0 + n_1$. In non-nested trial designs sampling randomized and non-randomized individuals with different sampling fractions induces a biased sampling model \cite{bickel1993efficient, breslow2000semi}, in the sense that the relative sample size of the trial and the emulation, $n_1/(n_0 + n_1)$, does not necessarily reflect the population proportion of trial participation \cite{dahabreh2020studydesigns, dahabreh2019generalizing}. Thus, the probability of trial participation $\Pr[S = 1]$ is not identifiable in non-nested trial designs, even though the densities $f(x|S = s)$ are identifiable for $s = 0, 1$ \cite{dahabreh2020studydesigns}.

Let $Y^a$ denote the counterfactual (potential) outcome under intervention that sets treatment $A$ to $a$ \cite{rubin1974, robins2000d}. We are interested in the conditional potential outcome mean in the trial $\E[Y^a | X, S = 1]$ and the emulation $\E[Y^a | X, S = 0]$, as well as the corresponding marginal (population-averaged) potential outcomes means $\E[Y^a | S = 1]$ and $\E[Y^a | S = 0]$. Under separate sampling into the trial and the emulation, it is not possible to identify $\E[Y^a]$ without additional information on how individuals are sampled into the two studies; such information is usually unavailable except when the trial is nested in a broader cohort of trial-eligible individuals (see references \cite{dahabreh2020studydesigns, dahabreh2020transportingStatMed} for details).

\section*{Identification in the trial and its emulation}\label{sec:identification}

\paragraph{Identifiability conditions:} To focus on benchmarking concepts, we consider a trial and its emulation in the idealized case of complete adherence to assigned treatment strategies, no loss-to-follow-up, and no measurement error (our approach can be generalized to address these complications).  

We now list sufficient conditions under which the potential outcome means conditional on $X$ under intervention that sets treatment $A$ to $a$ in the populations underlying the trial and its emulation are identifiable. Throughout this paper we take identifiability conditions that involve counterfactuals as primitive conditions (i.e., not derived) to focus on issues of design, analysis, and interpretation. We note, however, that the conditions can be derived from structural equation models for the data generating mechanism. In particular, we have recently discussed \cite{dahabreh2019identification, dahabreh2020benchmarking} how the distributional exchangeability conditions invoked throughout the paper can be derived from non-parametric structural equation models with a finest fully randomized causally interpretable structured tree graph error structure, represented using causal directed acyclic graphs and single-world intervention graphs \cite{richardson2013single} (see also references \cite{pearl2014, bareinboim2016causalfusion} for an alternative approach based on selection diagrams).

\vspace{0.1in}
\noindent
\emph{(i) No data source effects and consistency of potential outcomes:} if $A_i = a$, then $Y^{a}_i = Y_i$, for every treatment strategy $a$ and unit $i$, regardless of trial participation status.

\vspace{0.1in}
\noindent
\emph{(ii) Conditional exchangeability over $A$ in the trial:} for every treatment strategy $a$, $Y^a \indep A | (X, S = 1)$.

\vspace{0.1in}
\noindent
\emph{(iii) Positivity of treatment in the trial:} for every $x$ with $f(x , S = 1) \neq 0$ and every treatment strategy $a$, $\Pr[A = a | X = x, S = 1] > 0$. 

\vspace{0.1in}
\noindent
\emph{(iv) Conditional exchangeability over $A$ in the emulation:}  for every treatment strategy $a$, $Y^a \indep A | (X, S = 0)$.

\vspace{0.1in}
\noindent
\emph{(v) Positivity of treatment in the emulation:} for every $x$ with $f(x , S = 0) \neq 0$ and every treatment strategy $a$, $\Pr[A = a | X = x, S = 0] > 0$.

Condition \emph{(i)} contains an exclusion restriction assumption that participation in the trial does not affect the outcome except through treatment assignment \cite{dahabreh2019identification}, and may be violated when trial participation has direct effects on the outcome (e.g., through ancillary non-protocol directed treatments or Hawthorne effects). Condition \emph{(i)} also requires the interventions to assign treatment to be well-defined, and may be violated if there exist multiple outcome-relevant versions of treatment, especially if some versions are not available outside the experimental setting  \cite{vanderWeele2009, hernan2011compound, vanderweele2013causal}. Conditions \emph{(ii)} and \emph{(iii)} are expected to hold because of randomization. Note that if the trial is marginally randomized, the exchangeability condition $(Y^a, X) \indep A | S = 1$ will hold; this condition implies but is not implied by condition \emph{(ii)} and we use the weaker condition \emph{(ii)} in the remainder of the paper. Condition \emph{(iv)} is a strong untestable assumption about the emulation and needs to be evaluated in light of substantive background knowledge on a case-by-case basis; condition \emph{(v)} is in principle testable, but empirical assessment can be challenging when $X$ is high-dimensional \cite{petersen2012diagnosing}.

\paragraph{Identification:} In the trial, using conditions \emph{(i)}, \emph{(ii)}, and \emph{(iii)} \cite{hernan2020}, we obtain 
\begin{equation}\label{eq:cond_ident_S1}
    \E[Y^{a} | X, S = 1]  = \E[Y^{a} | X, S = 1 , A = a] = \E[Y | X, S = 1 , A = a],
\end{equation}
and, by the law of total expectation, $\E[Y^{a} | S = 1] = \E \big[ \E[Y | X, S = 1 , A = a] | S = 1 \big]$.

In the emulation, using conditions (\emph{i}), (\emph{iv}), and (\emph{v}) \cite{hernan2020}, we obtain 
\begin{equation}\label{eq:cond_ident_S0}
    \E[Y^{a} | X, S = 0] = \E[Y^{a} | X, S = 0 , A = a] = \E[Y | X, S = 0 , A = a].
\end{equation}
By the law of total expectation, the above result implies that the potential outcome mean under treatment assignment $a$ in the population underlying the emulation, $ \E[Y^{a} | S = 0]$, is identified by the observed data functional
\begin{equation}\label{eq:marg_ident_S0_obs}
    \phi(a) \equiv \E \big[ \E[Y | X, S = 0 , A = a] | S = 0 \big].
\end{equation}

Nothing in the above results suggests that the marginal or conditional potential outcome means in the two studies are related in any way and, in general, we should expect $ \E[Y^{a} | S = 1] \neq \E[Y^{a} | S = 0]$ and, consequently, $\E \big[ \E[Y | X, S = 1 , A = a] | S = 1 \big] \neq \E \big[ \E[Y | X, S = 0 , A = a] | S = 0 \big]$, even if conditions (\emph{i}) through (\emph{v}) hold. Up to this point in our exposition, the results simply pertain to the different populations underlying the trial and the emulation.

\section*{Benchmarking and joint analysis}\label{sec:benchmarking}

In order to connect the causal quantities in the trial and its emulation we can use concepts from the emerging literature on extending (generalizing or transporting) causal inferences from trials to a new population \cite{dahabreh2019commentaryonweiss, dahabreh2020benchmarking}. Suppose that we believe that selection into the trial is independent of potential outcomes, given baseline covariates, so that we can re-calibrate the trial results to the population underlying the emulation (similar ideas have appeared in references \cite{hartman2013, lodi2019effect, webster_clark2020}). More formally, suppose that we are also willing to assume the following two conditions: 

\vspace{0.1in}
\noindent
\emph{(vi) Conditional exchangeability over $S$:} for every treatment strategy $a$, $Y^a \indep S | X$. 

\vspace{0.1in}
\noindent
\emph{(vii) Positivity of participation:} for every $x$ with positive density $f(x) \neq 0$ and every $s$, $\Pr[S = s | X = x] > 0$.

Note that condition (\emph{vi}) essentially requires the baseline covariates $X$ to be adequate for addressing selective participation in the trial \cite{steg2007external}, in addition to being sufficient for addressing confounding in the emulation. For simplicity, here, we assume that the set of baseline covariates in conditions  (\emph{vi}) and  (\emph{vii}) is the same as in conditions  (\emph{iv}) and  (\emph{v}). In the Appendix, we consider the slightly more complicated case where a different set of covariates is needed for for each pair of conditions.

\subsection*{Benchmarking}

We will now argue that conditions \emph{(vi)} and \emph{(vii)}, when combined with conditions \emph{(i)} through \emph{(v)}, impose restrictions on the law of the observed data and allow the formal benchmarking of observational analysis methods against trials.

\paragraph{Restrictions on the law of the observed data:} Under conditions (\emph{vi}) and (\emph{vii}) we have $\E[Y^{a} | X, S = 1] = \E[Y^{a} | X, S = 0]$, that is, the far left-hand-sides in equations (\ref{eq:cond_ident_S1}) and (\ref{eq:cond_ident_S0}) are equal. Combined with conditions (\emph{i}) through (\emph{v}), this result implies that the far right-hand-sides of equations (\ref{eq:cond_ident_S1}) and (\ref{eq:cond_ident_S0}) are also equal, $\E[Y | X, S = 1 , A = a] = \E[Y | X, S = 0 , A = a]$. Informally, \emph{if} the conditions needed for \emph{both} the trial and emulation results to have a causal interpretation, \emph{and if} results are transportable between the populations underlying the two studies, \emph{then} the conditional (observed) outcome mean among randomized individuals assigned to treatment $a$ in the trial has to equal the conditional outcome mean among individuals who chose to receive treatment $a$ in the emulation. 

Because the equality $\E[Y | X, S = 1 , A = a] = \E[Y | X, S = 0 , A = a]$ only involves the observed variables, investigators can examine whether it is compatible with the data. Investigators can use various qualitative or quantitative approaches to examine whether this implication of the assumptions is compatible with the data \cite{delgado1993testing, neumeyer2003nonparametric, racine2006testing}. A detailed discussion of such approaches is beyond the scope of this paper; however, we advise against over-reliance on formal statistical testing of the observed data implications of the identifiability conditions. We also note that under our identifiability conditions, when the outcome is not binary, it is possible to deduce restrictions on the law of the observed data that are stronger than the equality of the two conditional expectations. Specifically, in the Appendix, we show that exchangeability conditions (\emph{ii}), (\emph{iv}), and (\emph{vi}), together with the consistency condition (\emph{i}), imply that $Y \indep S | (X, A = a)$ for every treatment $a$. Thus, the distribution of the observed outcome $Y$ is independent of $S$, conditional on $X$, within each level of treatment $A$. In other words, the implications of the conditions extend beyond mean independence, to independence of the entire distribution of the outcome from the indicator for trial participation, conditional on baseline covariates and within levels of treatment. Here, we only consider mean independence.

\paragraph{The formal logic of benchmarking:} Under conditions (\emph{i}), (\emph{ii}), (\emph{iii}), (\emph{vi}), and (\emph{vii}) the potential outcome mean in the population underlying the emulation, $  \E[Y^{a} | S = 0]$, can be identified by the observed data functional
\begin{equation}\label{eq:marg_ident_S0_tran}
    \chi(a) \equiv \E \big[ \E[Y | X, S = 1 , A = a] | S = 0 \big].
\end{equation}

Note that the right-hand-sides of displays (\ref{eq:marg_ident_S0_obs}) and (\ref{eq:marg_ident_S0_tran}) are different and the equality in each of these displays depends on different sets of conditions -- only condition (\emph{i}) is needed for both results. 

These identification results provide the formal logic of benchmarking: under conditions (\emph{i}), (\emph{iv}), and (\emph{v}), $\phi(a)$ can be given a causal interpretation as $\E[Y^a | S = 0]$; under conditions (\emph{i}), (\emph{ii}), (\emph{iii}), (\emph{vi}), and (\emph{vii}), a different quantity, $\chi(a)$, has the same interpretation. Thus, to the extent that the data can be used to estimate both $\phi(a)$ and $\chi(a)$, when the estimates are substantially different (i.e., beyond what would be expected by sampling variability) we might reasonably conclude that at least one of the conditions (\emph{i}) through (\emph{vii}) are violated. 

Conversely, if estimates of $\chi(a)$ are approximately equal to those of $\phi(a)$ it is reasonable to think that the conditions are not grossly violated. This is a sensible inference, though not logically necessary: It is possible to have $\phi(a)\approx \chi(a)$ when $\E[Y^a | X, S = 0, A = a] = \E[Y^a | X, S = 0]$ and $\E[Y^a | X, S = 1] = \E[Y^a | X, S = 0]$ hold, even if the distributional exchangeability conditions are violated. Furthermore, even if $\E[Y^a | X, S = 0, A = a] \neq \E[Y^a | X, S = 0]$ or $\E[Y^a | X, S = 1] \neq \E[Y^a | X, S = 0]$, it is still possible to have $\chi(a) \approx \phi(a)$ because these conditional expectations are averaged over the covariate distribution of the population underlying the emulation (so differences can ``cancel-out''). Furthermore, it is possible, though unlikely, that multiple conditions fail in a way that still leads to $\chi(a) \approx \phi(a)$.

In the section of this paper on estimation, we show how data from a trial and its emulation can be used to efficiently and robustly estimate $\phi(a)$ and $\chi(a)$ and how to use the estimates for benchmarking purposes.

\paragraph{The heuristic logic of benchmarking, revisited:} We now revisit the logic of benchmarking, sketched in the Introduction, in view of our formal results. To the extent that condition (\emph{vi}) is deemed plausible, the leading (but, as discussed in the preceding section, not the only possible) explanation for $\chi(a)$ to be different from $\phi(a)$ is violation of condition (\emph{iv}). If estimates of these two quantities are similar we might believe that in some sense the methods in the observational emulation were successful. Such informal benchmarking of observational analyses is methodologically interesting \cite{dahabreh2014can}, as evidenced by numerous meta-epidemiological studies comparing independently conducted trial and observational analyses (e.g., \cite{concato2000randomized, benson2000comparison, dahabreh2012observational, kitsios2015can, lonjon2014comparison}) and systematic attempts to emulate trials using observational data \cite{franklin2021emulating}. It is also substantively important because it allows investigators to develop trust in observational analyses in order to use them in contexts where trial data are not available. We note, however, that if condition (\emph{vi}) is violated, as would be the case, for instance, if the populations underlying the trial and its emulation differ with respect to an unmeasured important outcome predictor, estimates of $\chi(a)$ and $\phi(a)$ may disagree, even in the absence of unmeasured confounding in the emulation.

Because of the multitude of assumptions involved in benchmarking, disagreement between a trial and its emulation will not necessarily mean that the emulation produced estimates that do not have a causal interpretation. To see this suppose that conditions (\emph{i}) through (\emph{iii}), (\emph{v}), and (\emph{vii}) all hold. In that situation, the only assumptions in question pertain to the critical exchangeability conditions (\emph{iv}) $Y^a \independent A | (X, S = 0)$ (i.e., no confounding in the population underlying the emulation) and (\emph{vii}) $Y^a \independent S | X$ (i.e., transportability among the two populations). In Table \ref{table:interpretation} we summarize how the possible truth values of these conditions relate to the observed data independence condition $Y \indep S | (X, A = a)$, and for interpreting the results of the emulation and benchmarking.

\vspace{0.5in}
\renewcommand{\arraystretch}{1.2}
\begin{table}[ht]
\caption{Truth values of conditions (\emph{iv}) $Y^a \independent A | (X, S = 0)$ and (\emph{vi}) $Y^a \independent S | X$ and their implications for the observed data independence $Y \indep S | (X, A = a)$ and for interpreting benchmarking results, provided conditions (\emph{i}) through (\emph{iii}), (\emph{v}), and (\emph{vii}) hold.}\label{table:interpretation}
\footnotesize
\begin{tabular}{|c|c|c|c|c|c|}
\hline
\multirow{2}{*}{($\emph{iv}$)} & \multirow{2}{*}{($\emph{vi}$)} & \multirow{2}{*}{$Y \indep S | (X, A = a)$} & \multicolumn{3}{c|}{Interpretation}                                                                                                                                                                                                                                                            \\ \cline{4-6} 
                               &                                &                                            & \begin{tabular}[c]{@{}c@{}}Do emulation estimates \\ have a causal \\ interpetation?\end{tabular} & \begin{tabular}[c]{@{}c@{}}Is benchmarking \\ possible?\end{tabular} & \begin{tabular}[c]{@{}c@{}}Will benchmarking show \\ agreement between the\\  trial and its emulation?\end{tabular} \\ \hline
T                              & T                              & has to hold                                          & yes                                                                                               & yes                                                                  & yes                                                                                                                 \\ \hline
F                              & T                              & does not have to hold                                          & no                                                                                                & yes                                                                  & no                                                                                                                  \\ \hline
T                              & F                              & does not have to hold                                          & yes                                                                                               & no                                                                   & NA                                                                                                                  \\ \hline
F                              & F                              & does not have to hold                                          & no                                                                                                & no                                                                   & NA                                                                                                                  \\ \hline
\end{tabular}
\caption*{T = true; F = false; NA = not applicable.}
\end{table}

Let us now discuss the practical implications of this table for benchmarking efforts. Suppose the data are consistent with the independence condition $Y \indep S | (X, A = a)$ (first row of the table). Then our (subjective) belief in the following three hypotheses would be strengthened: (1) conditions (\emph{iv}) and (\emph{vi}) both hold; (2) the emulation estimates have a causal interpretation; (3) the treatment effect is the same in the emulation as in transportability analyses from the trial to the population underlying the emulation. We are not aware of any systematic empirical assessments of independence conditions such as $Y \indep S | (X, A = a)$ in ongoing efforts to emulate trials using observational analyses; such assessments may be an interesting extension of ongoing research.   

Next, suppose that the data are inconsistent with the independence condition $Y \indep S | (X, A = a)$. The leading explanation for this finding would be a violation of condition (\emph{iv}), (\emph{vi}), or both (bottom three rows of the table). Unfortunately, the data cannot distinguish between these three possibilities. In other words, we cannot tell whether the emulation estimates do not have a causal interpretation and this manifests as a benchmarking discrepancy (second row: confounding in the emulation; exchangeable populations underlying the trial and the emulation); the emulation estimates have a causal interpretation but formal benchmarking is not possible (third row: no confounding in the emulation, non-exchangeable populations underlying the trial and the emulation); or the emulation estimates do not have a causal interpretation and formal benchmarking is impossible (fourth row: confounding in the emulation, non-exchangeable populations underlying the trial and the emulation). This is a somewhat sobering result for the interpretation of benchmarking attempts that uncover a discrepancy between (1) estimates based on extending inferences from the trial to the population underlying the emulation and (2) the emulation itself: the discrepancy may reflect a failure of the emulation (e.g., due to unmeasured confounding) or inability to benchmark (e.g., due to selective participation into the trial), but we cannot use the data to decide which is the case.

\subsection*{Joint analysis}\label{sec:joint_analysis}

If conditions \emph{(i)} through \emph{(vii)} all hold, then it is clear that we can identify $\E[Y^a | S = 0]$ using either $\phi(a)$ or $\chi(a)$. We now argue that a third identification result is also available. As we saw above, when conditions (\emph{i}) through (\emph{vii}) hold, $$\E[Y | X, S = 1 , A = a] = \E[Y | X, S = 0 , A = a] = \E[Y | X, A = a].$$ Consequently, if conditions \emph{(i)} through \emph{(vii)} hold, we can identify $\E[Y^a | S = 0 ]$ by the observed data functional $$\psi(a) \equiv  \E \big[ \E[Y | X, A = a] \big | S = 0 \big].$$

This result suggests that we can completely pool the data from the trial and its emulation when modeling the conditional outcome mean (because $\E[Y | X, A = a]$ in the formula does not condition on trial participation status $S$) but still obtain inferences about the population underlying the emulation, which may be more representative of routine clinical practice (by standardizing the conditional outcome mean over the distribution of covariates in $S = 0$). Thus, $\psi(a)$ provides a way to combine data from the trial and its emulation in a joint analysis that may have a more natural causal interpretation than standard meta-analyses combining estimates from trial and observational analyses \cite{dahabreh2020toward}.

\section*{Estimation and inference}\label{sec:estimation}

In this Section we informally describe some results about estimation and inference for benchmarking and joint analyses using a trial and its emulation. The focus here is on the intuition behind our results; additional details are provided in the Appendix. 

\subsection*{Benchmarking}

As noted, benchmarking involves a comparison of $\phi(a)$ versus $\chi(a)$. Such a comparison should ideally use efficient and robust methods to statistically compare these quantities using the data.

Because $\phi(a)$ is a version of the well-known g-formula identification result for observational analyses, we can use a well-known doubly robust estimator to estimate it \cite{tsiatis2007}; we give a formula for this estimator in the Appendix and denote it as $\widehat \phi(a)$. The estimator depends on estimating the probability of treatment in the emulation, $\Pr[A = a | X, S = 0]$, and the expectation of the outcome in each treatment group in the emulation, $\E[Y | X, S = 0, A = a]$. The estimator $\widehat \phi(a)$ is doubly robust \cite{bang2005} in the sense that it remains consistent when either the estimator for  $\Pr[A = a | X, S = 0]$ or the estimator for  $\E[Y | X, S = 0, A = a]$ is consistent (but not necessarily both). 

We propose to estimate $\chi(a)$ using an estimator for transporting inferences from the trial to a target population \cite{dahabreh2020transportingStatMed}; we also give a formula for this estimator in the Appendix and denote it as $ \widehat \chi(a)$. This estimator depends on estimating the probability of participation in the trial, $\Pr[S = 1 | X]$; the expectation of the outcome in each treatment group in the trial, $\E[Y | X, S = 1, A = a]$; and the probability of treatment in the trial, $\Pr[A = a | X, S = 1]$. This estimator is doubly robust in the sense that it remains consistent when either the estimator for $\Pr[S = 1 | X]$ or the estimator for $\E[Y | X, S = 1, A = a]$ is consistent (but not necessarily both); the probability of treatment among trial participants is known and thus $\Pr[A = a | X, S = 1]$ can always be consistently estimated.

The benchmarking contrast $\phi(a) - \chi(a)$ can be estimated by taking the difference $\widehat \phi(a) - \widehat \chi(a)$ as the basis for formal (statistical) comparisons between the trial and its emulation. Confidence intervals for the component quantities or for the difference contrast can be obtained using the usual sandwich methods \cite{boos2013essential}; the variance of the influence curves (based on the empirical analogs of the influence functions given in the Appendix; or the bootstrap \cite{efron1994introduction}.

\subsection*{Joint analysis}

As noted above, the estimators for benchmarking are based on previously obtained results for observational analyses with no unmeasured confounding or for transportability analyses from a trial to a new target population. In the Appendix, we propose a novel estimator for $\psi(a)$, which we denote as $\widehat \psi(a)$. This estimator depends on models for the probability of trial participation, $\Pr[S = 1 | X]$; the probability of treatment in the pooled data, $\Pr[A = a | X]$; and the expectation of the outcome in each treatment group, $\E[Y | X, A = a]$ in the pooled data. This estimator is also doubly robust, in the sense that it is consistent when either the pair of estimators for $\Pr[S = 1 | X]$ and  $\Pr[A = a | X]$ are consistent, or when the estimator for $\E[Y | X, A = a]$ is consistent.

When conditions (\emph{i}) through (\emph{vii}) hold, all three estimators -- $\widehat \phi(a), \widehat \chi(a), \widehat \psi(a)$ -- are potentially useful for estimating potential outcome means in the target population. If all models are correctly specified (and converge to the ``true'' model at a fast enough rate), then all estimators are consistent and asymptotically normal; if models are misspecified all three estimators enjoy certain robustness properties (see Appendix). A reasonable way to choose among estimators, then, is to use the one with the lowest large-sample variance. In the Appendix, we give expressions for the large-sample variance bounds of regular estimators for each of $\phi(a)$, $\chi(a)$, and $\psi(a)$, and argue that the corresponding estimators attain them when models are correctly specified. We also show that the asymptotic variance bound for $\psi(a)$ is smaller than or equal to the asymptotic variance bounds of $\phi(a)$ and $\chi(a)$. This should make intuitive sense: $\psi(a)$, by taking advantage of the complete set of conditions (\emph{i}) through (\emph{vii}), allows analysts to use the covariate, treatment, and outcome data from all observations to estimate the necessary working models; when these models are correctly specified, $\widehat \psi(a)$ makes the most efficient use of the data. In the Appendix, we discuss some additional technical issues that arise when estimating the various models required for $\widehat \phi(a), \widehat \chi(a), \widehat \psi(a)$.

\section*{Discussion}

We have provided a formal description of analyses that synthesize randomized and observational data in order to benchmark methods for analyzing the observational data and, when appropriate, perform joint analyses. Our description combines ideas from the large literature on conducting observational analyses that emulate target trials \cite{hernan2016} with ideas from the emerging literature on analyses extending inferences from a trial to a target population \cite{dahabreh2019commentaryonweiss}, and provides a basic conceptual framework for thinking about analyses that combine aspects of both (e.g., \cite{lodi2019effect, webster_clark2020}). Our framework has some parallels with an independently proposed framework for replication studies and within-study comparisons \cite{wong2018replication, wong2018designs}, though the study designs and methods we describe are different.

Using trial and observational data to benchmark methods has a long history, including seminal works in econometrics \cite{lalonde1986evaluating, fraker1987adequacy}. Using data from a trial nested in a cohort of trial eligible individuals, including those who were not randomized (i.e., a nested trial design \cite{dahabreh2020studydesigns}) in a joint analysis also has a long history (e.g., see \cite{olschewski1992} and the more modern treatment, similar to our approach, in \cite{lu2019causal}). Our contribution here is the consideration of these issues when using a modern causal and statistical approach in the much more common case where the trial and its emulation are conducted independently (i.e., non-nested trial design \cite{dahabreh2020studydesigns}). The non-nested design is more common for benchmarking because, unlike the nested design, it does not require prospective nesting of the trial in a well-defined cohort or the ability to retrospectively link data from trial participants to data sampled from a population that also includes non-randomized individuals.

To simplify exposition, we worked in a simple setting with complete adherence to treatment and outcomes measured at the end of follow-up for all individuals. This setting is adequate for illustrating the concepts of benchmarking and emulation; results from the simple setting can be extended to more realistic settings with non-adherence, longitudinal or failure time outcomes, and loss-to-followup \cite{dahabreh2019identification}, as well as to the consideration of multiple studies jointly \cite{dahabreh2020toward}. Thus, analogous identification results should be fairly easy to obtain in more realistic settings. Nevertheless, we expect that estimation using real data in applied analyses will prove more challenging and that practical experience with such analyses will take time to accumulate. We hope that the results herein provide a reasonable conceptual basis for developing a broader framework and using the methods in practice.

Our results suggest that, even in a simplified setting, benchmarking combines causal assumptions, data analysis methods, and substantive knowledge to build qualitative arguments about the validity of observational analysis methods. Different investigators will find such arguments convincing to varying degrees. Some, including ourselves, will be convinced when (1) the observational analysis is explicitly designed to emulate a target trial with a protocol sufficiently similar to that of the index trial; (2) the emulation is based on sufficiently rich observational data; (3) both the index trial and the emulation have the same causal estimand and are analyzed using comparable (and appropriate) methods; and (4) the results of the observational emulation and the index trial are similar. Naturally, in each clinical domain, different investigators will require different kinds of evidence (e.g., in different populations, or conducted concurrently with trials) and different amounts of evidence (e.g., more emulations, by independent teams, using different data sources) to increase trust in observational analyses.

\clearpage
\bibliographystyle{unsrt}
\bibliography{bibliography}{}


\ddmmyyyydate 
\newtimeformat{24h60m60s}{\twodigit{\THEHOUR}.\twodigit{\THEMINUTE}.32}
\settimeformat{24h60m60s}
\begin{center}
\vspace{\fill}\ \newline
\textcolor{black}{{\tiny $ $benchmarking\_joint\_analysis, $ $ }
{\tiny $ $Date: \today~~ \currenttime $ $ }
{\tiny $ $Revision: \paperversionmajor.\paperversionminor $ $ }}
\end{center}

\end{document}